%% file: bare_jrnl.tex
\documentclass[10pt, a4paper]{IEEEtran}

\usepackage{amsmath,amssymb,amsthm}
\usepackage{algorithmic}
\usepackage{array}
\usepackage{fixltx2e}
\usepackage{stfloats}
\usepackage[utf8]{inputenc}	
\usepackage[british]{babel}
\usepackage{csquotes}
\usepackage[pdftex]{graphicx}
\graphicspath{ {image/} }
\usepackage{amssymb}
\usepackage{mathtools}
\usepackage{todonotes}
\usepackage{authblk}

\begin{document}
\title{Experimental Validation of Time Reversal Multiple Access for UWB Wireless Communications Centered at the $273$ GHz Frequency}

\author[1]{Ali Mokh}
\author[1]{Julien de Rosny}
\author[2]{George C. Alexandropoulos}
\author[3]{\\Mohamed Kamoun}
\author[1]{Abdelwaheb Ourir}
\author[3]{Ramin Khayatzadeh}
\author[1]{Arnaud Tourin}
\author[1]{Mathias Fink}

\affil[1]{ESPCI Paris, PSL Research University, CNRS, Institut Langevin, France}
\affil[2]{Department of Informatics and Telecommunications,
National and Kapodistrian University of Athens, Greece}
\affil[3]{Mathematical and Algorithmic Science Lab, Paris Research Center, Huawei Technologies France}
\affil[ ]{emails: firstname.lastname@espci.fr, alexandg@di.uoa.gr, firstname.lastname@huawei.com\vspace{-0.6cm}}



\maketitle

\begin{abstract}
Ultra-high data rates with low-power consumption wireless communications and low-complexity receivers is one of the key requirements for the next $6$-th Generation (6G) of communication networks. Sub-Terahertz (SubTHz) frequency bands can support Ultra-WideBand (UWB) communications and can thus offer unprecedented increase in the wireless network capacity, rendering those bands and relevant research a strong candidate technology for 6G. However, in contrast to millimeter-wave communications, the technological advances in subTHz transceivers are not yet completely mature. In this paper, we focus on the Time Reversal (TR) precoding scheme, which is a computationally simple and robust technique capable of focusing UWB waveforms on both time and space, by exploiting channel diversity when available. We design a novel experimental setup, including a THz link inside a waveguide and operating at the carrier frequency $273.6$ GHz with $2$ GHz transmission bandwidth, and deploy it to perform multi-user communications with simple modulation. Our results demonstrate large communication rates with only $3$ mm space separation of the receiving antennas.
\end{abstract}


\IEEEpeerreviewmaketitle
\section{Introduction}
The technologies for the Internet of Things (IoT) offer advanced services in different domains, such as medicine, power distribution, connected smart cities, smart homes, and farming/agriculture. Some of their key requirements are the support of massive deployment, scalability, high efficiency, low power consumption, and intermittent connectivity. 

An important part of various IoT applications is the wireless communication network which needs to support low-power devices equipped with ultra-low complexity transceivers \cite{georgiou2017iot,luo2009analysis,gray2015power}. Various communication strategies have been proposed to limit the energy consumption of the communication modem of IoT devices, both from the software and hardware perspectives \cite{gray2015power,henkel2017ultra}. In particular, UWB waveforms have ushered a new era in short-range communications for wireless sensor networks, by enabling both robust communications and accurate ranging capabilities \cite{zhang2009uwb}.
To support large bandwidth for multiple devices, the recent initial discussions indicate that frequencies in the range of sub-TeraHertz (subTHz) and above will be considered for $6$-th Generation (6G) wireless network. Those frequencies include large chunks of available bands that are suitable to satisfy the aforementioned requirements \cite{saad2019vision,tariq2020speculative}. More specifically, it appears that 6G will utilize spectra beyond $140$ GHz with particular applications in very short-range communication or ‘whisper radio’ \cite{xing2018propagation}. One central issue facing Ultra-WideBand (UWB) waveforms for low-complexity devices is the efficient energy collection that is dispersed in rich scattering signal propagation environments. 

The Time Reversal (TR) technique is experimentally proven to exploit rich multipath channels to focus UWB pulses in both the time and space dimensions \cite{lerosey2004time}, while shifting the signal processing complexity to the transmitter side and keeping the reception simple. In \cite{Qiu_JSAC_2006}, TR was considered in UWB multi-user wireless communications. The signal-to-noise-ratio improvement, thanks to TR's temporal focusing capability, has been experimentally verified with a multi-antenna transmitter \cite{TR_MISO_AWPL_2006, mokh2021indoor,mokh_WCNC}. The concept of TR Division Multiple Access (TRDMA) was introduced in \cite{TR_MIMO_MA_2012} for multi-user communication. Very recently \cite{mokhTHz}, the spatio-temporal focusing capability of TR in the subTHz domain was evaluated, and the experimentally measured channel were used to simulate data transmission with the spatial modulation technique.
 
In this paper, we experimentally evaluate the capability of the TR technique in the subTHz domain to offer multiple access for low-complexity devices at high data rates. We present the designed experimental setup to perform channel estimation and TR-based transmission at the $273.6$ GHz carrier frequency and with a bandwidth up to $2$ GHz. The link between the transmitter and the receiver was a waveguide with a tilted incident angle to create a multipath wireless channel. We selected two different positions for the receiver, to emulate two different receivers, and we transmitted binary data using a modulation technique that depends on the position of the peak of the received signal. The detection was performed via two incoherent receivers with a simple power detector. 

The rest of the paper is organized as follows. In Section~II, the principle of TRDMA is described, while section~III includes the subTHz experimental setup to estimate the channels and to measure the TR signals. In section~IV, the experimental results for data transmission with different data rates are presented. Finally, section~V concludes the paper.

\section{Time Reversal Division Multiple Access}
We consider a multi-user wireless communication system comprising one transmitter that is equipped with a single antenna element, which wishes to communicate in the downlink direction with $N$ single-antenna receivers. We represent by $h_{i}[k]$ the baseband Channel Impulse Response (CIR) at discrete time $k$ between the transmitting and the $i$-th ($i=1,2,\ldots,N$) receiving antennas, i.e., at the base station and the $i$-th user, respectively. The TR precoding technique utilizes the time-reversed CIR, i.e., $h^*_{i}[L-k]$, to focus the transmitted information-bearing electromagnetic field on the $i$-th user location. As a consequence, the signal sent by the transmit antenna at discrete time $k$, to focus the baseband message $x_i$ on each individual receiving antenna, is given as follows:
\begin{align}
     s[k]=\sum_{i=1}^N  \frac{\sum_{l=1}^L x_i[l] h_{i}^*[L+l-k]}{\sqrt{\sum_{l=1}^L \lvert h_{i}[l] \rvert^2}}, 
     \label{eq:yi}
\end{align}
where it is assumed that the CIR is composed of $L$ significant taps. The normalization in this expression ensures that the power emitted toward each user is the same. 

We consider pulses to estimate the CIR for each user at the transmitter in the uplink direction, and then perform TR-based precoding. We let $T_p$ denote the pulse duration, which is related to the system bandwidth $B$ as $T_p=1/B$. Let also $D$ represent the rate backoff factor, which is introduced to match the symbol rate with the pulse duration, yielding the symbol duration $T_s=D T_p$. Then, the signal received by each user in the downlink is sampled every $T_s$ seconds, i.e., at each $t=kDT_p$ with $k=1,2,\dots$, in order to detect each symbol $x_i[k]$. The mathematical expression for the baseband received signal at each $j$-th ($j=1,2,\ldots,N$) user is given by:
\begin{equation}
   y_j[k]= \sum_{l=1}^L \sum_{i=1}^N x_i[l] R_{j,i}[L+l-kD]+n_j[k].
   \label{eq:TR}
\end{equation}
where the expression of the correlation function $R_{j,i}[k]$ between the CIRs of users $j$ and $i$ is defined as 
\begin{equation}
 	R_{j,i}[k] \triangleq \frac{\sum_{k'=-\infty}^{+\infty} h_{i}^*[k-k'] h_{j}[k']}{\sqrt{\sum_{l=1}^L \lvert h_{i}[l] \rvert^2}},
   \label{eq1}
\end{equation}
and $n_j[k]$ represents the zero-mean additive white Gaussian noise with standard deviation $\sigma$ at user $j$. 

The TR-based precoder aims to focus the transmitted signal in both time and space, by applying the time-reversed CIR, resulting from the intended receiver in the uplink phase, as the precoding scheme \cite{fink2001acoustic}. This means that a signal peak given by the CIR's auto-correlation function, i.e., at $R_{i,i}[0]=\sqrt{\sum_{l=1}^L \lvert h_{i,m}[l] \rvert^2}$, appears at user $i$; at this time and location, $ML$ signals will add up coherently. However, the imperfect spatiotemporal focusing of TR scheme can produce the following two forms of interference:
\begin{itemize}
    \item \textit{Inter-Symbol Interference (ISI):} this results from the signals that add up incoherently at each same antenna $i$ in time instants that are different from $0$, i.e., $R_{i,i}[k]$ at any $k\neq0$.
    \item \textit{Inter-User Interference (IUI):} this results from all the signals in the auto-correlation function for $i\neq j$.
\end{itemize}

\input{mmWaveTHz}

\bibliographystyle{IEEEtran}
\bibliography{IEEEabrv,ref}


\end{document}

%% file: mmWaveTHz.tex
\section{Experimentation At SubTHz Frequencies}
In this section, we experimentally investigate the TR precoding scheme in the subTHz frequency band. It is noted that, in such frequencies, the attenuation of the transmitted signal becomes very high, and it is difficult to have a multipath channel, even in indoor applications. We next present the designed experimental setup, resulting in multipath signal propagation, together with the application of TR-based precoding. Experimental results for the spatiotemporal focusing capability of TR are presented and discussed.

\subsection{Experimental Setup}
\begin{figure*}
\centering
\includegraphics[width=1 \linewidth]{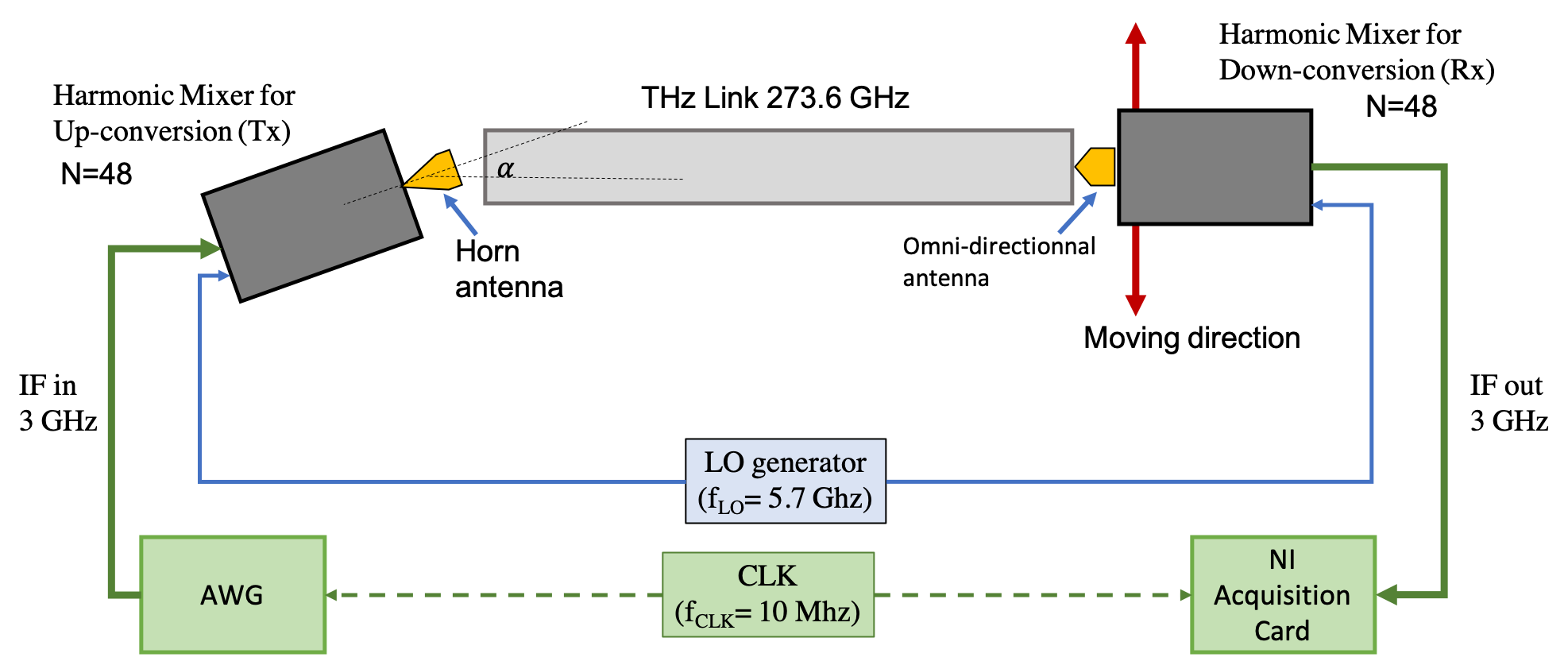}
\caption{The designed experimental setup at the central frequency $f_{c}=273.6$ GHz deploying an aluminum tube waveguide for multipath propagation.}
\label{setup}
\end{figure*}
Our experimental setup at the central frequency $f_{c}=273.6$ GHz using an aluminum tube as a waveguide to create a multipath channel is depicted in Fig$.$~\ref{setup}. The diameter of the cylindrical waveguide is equal to $7$ mm, i.e., about $10$ times the wavelength, and its length is $1$ m. The setup deploys $2$ Spectrum/Signal Analyzer eXtension modules (SAXs) by Virginia Diodes, Inc., where one of them is used at the Transmitter (Tx) side to upconvert the Intermediate Frequency (IF) signal with frequency $3$ GHz, generated by an I/Q modulation generator, to the carrier frequency $273.6$ GHz. This frequency is obtained thanks to the multiplication by a factor $N=48$ of the frequency of the considered Local Oscillator (LO) working at $5.7$ GHz. For the transmitter, we use a diagonal horn antenna provided by Virginia Diodes, Inc., while the transmitted electromagnetic field is received by an omnidirectional antenna. On the Receiver side (Rx), the same LO with the transmitted is used to downconvert the high-frequency signal to the IF frequency. Then, the IF signal is sampled by a waveform digitizer (an NI acquisition card). The modulation generator (i.e., Arbitrary Waveform Generator (AWG)) and the waveform digitizer are synchronized by a $10$ MHz reference CloCK (CLK), whereas the bandwidth is limited to $2$ GHz by the modulation generator. Instead of directly dealing with IF signals, for convenience, the recorded signals are numerically downconverted into in-phase and quadrature-phase pairs of signals. Each time step, that lasts the inverse of the I/Q modulator bandwidth, is called a tap. To probe the focal spot at the waveguide receive extremity, the dipole antenna is mounted on a linear stepper motor stage.

\subsection{CIR Estimation and TR-Based Focusing}
The estimation of the CIR in the uplink direction and the TR-based precoding was performed as follows. First, a chirp signal spanning the range $[f_c-B/2, f_c+B/2]$ was transmitted by the transmit antenna. Synchronously, the signal was probed by the receive antenna. A desktop computer was used to perform the CIR estimation by correlating the recorded signal with the chirp one. It is noted that, even if TR generates high amplitude pulses, the resulting gain is usually not sufficient to directly probe the pulse with a receiving antenna. For this reason, before transmission, the time-flipped CIR was also convoluted with the aforementioned chirp signal. Similar to the CIR estimation, the TR-based receive signal was reconstructed in the desktop computer via cross-correlation. 

\begin{figure}
\centering
\includegraphics[width=\columnwidth]{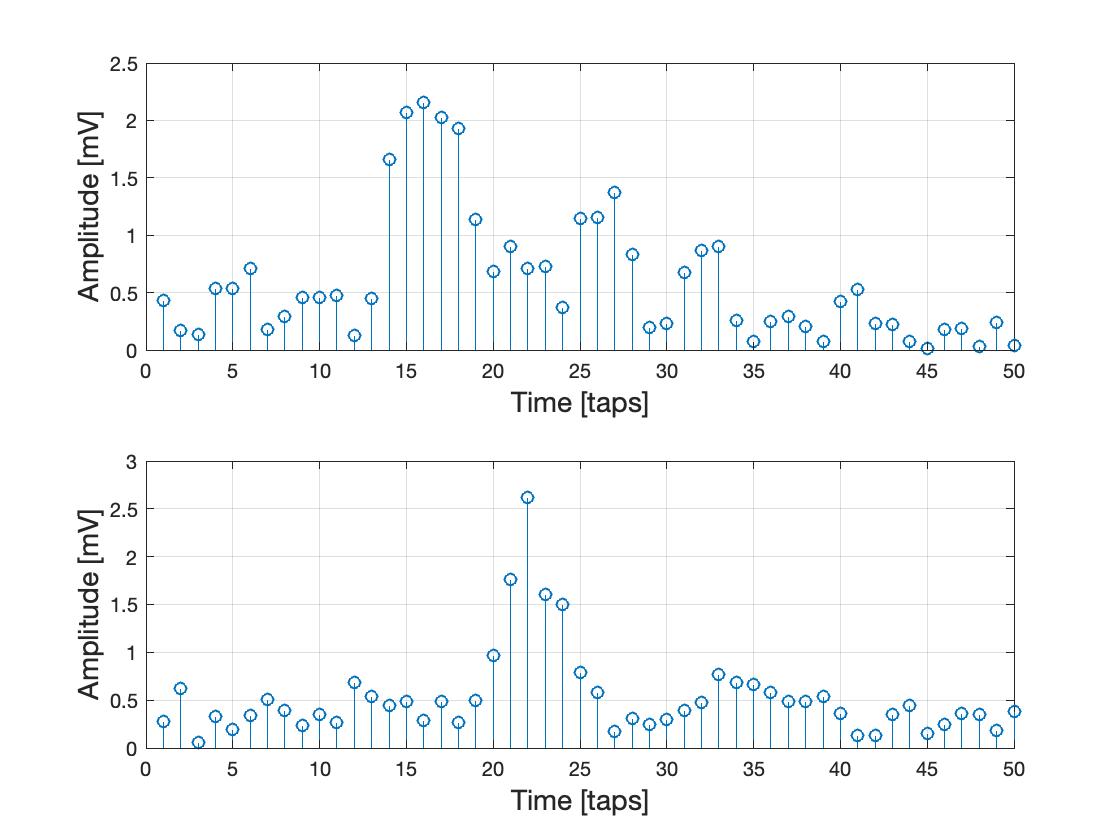}
\caption{The CIR estimation (up) and the received TR-precoded signal (bottom) spanning the bandwidth $B=2$ GHz.}
\label{CIR}
\end{figure}

The results for CIR estimation and the TR-precoded received signal for the considered $2$ GHz bandwidth are demonstrated in Fig$.$~\ref{CIR}. As shown, thanks to the deployed waveguide, the rays coming from the transmitter, which is inclined with the angle $\alpha$ from the straight line direction of the tube, exhibited different path distances because of their reflections inside the tube. This setting created a multipath channel that was probed at the receiver side, and its CIR is depicted at the top subfigure of Fig$.$~\ref{CIR}. The bottom subfigure illustrates the received TR-precoded signal that is focused in the time instant $22$ which has the largest amplitude.

\subsection{Intended vs Unintended Focusing}
\begin{figure}
\centering
\includegraphics[width=1\columnwidth]{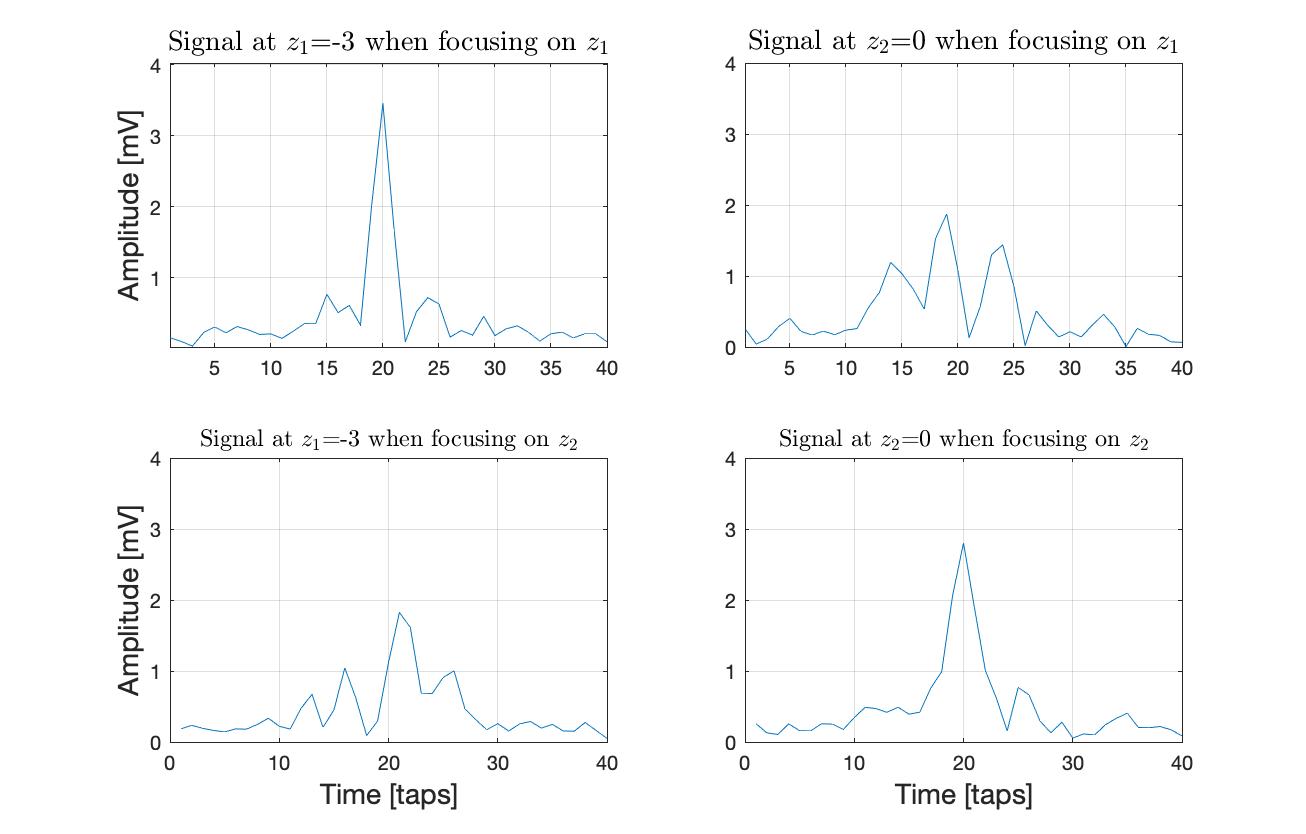}
\caption{Baseband received signals at the two spatial positions $z_1=-3$ and $z_2=0$ when applying TR to focus towards $z_1$ (up) or $z_2$ (bottom).}
\label{MUSignal}
\end{figure}

Using the transmission bandwidth $B=2$ GHz, we measured the CIR for different positions of the receiver antenna spaced by $1$ mm, and specifically between $x=-3$ mm to $3$ mm, where $x=0$ is the position facing the center of the tube's output. We selected two receiver positions where the CIRs were less correlated, namely $z_1=-3$ and $z_2=0$, and applied the TR precoding. Figure~\ref{MUSignal} shows the baseband received signals at the two spatial positions $z_1$ and $z_2$ when applying TR to focus towards $z_1$ or $z_2$, i.e., the focusing signal and the IUI when targeting one of the two positions. A peak signal localized in time appears at the targeted position for both considered cases. In addition, the figure illustrates the capability of TR to focus the transmitted signal in time and space at an intended receive position, while keeping it in low level in unintended ones.

To investigate the capability of TR to transmit data in multi-access scenarios, we use the aforedescribed scenario to modulate binary data with the focusing signal, using a modulation technique that is explained in the following section. 

\section{TRDMA with Pulse Position Modulation}
In this section, we describe the experimented data communication system using Pulse Position Modulation (PPM) \cite{PPM} and present experimental measurements for the capability of TR to offer multi-access communications. 

\subsection{Data Communication Model and Bit-Rate Results}
We have exploited the channel spatial diversity, that resulted from the designed experimental setup in Section~III.A, to transmit via TR-based precoding two data streams at two mm-level spaced positions at the same time and frequency. As explained in the previous section, to accomplish this, we convoluted the reversed CIR (for TR precoding) by a chirp to increase its energy, since we were limited by the maximum power of the transmitter, and hence, the duration of each symbol was related to the chirp duration. For this reason, we decided to use PPM to code the binary information, according to which, the position of the maximum amplitude of the received signal, or the maximum of the auto-correlation function in \eqref{eq1}, codes the information signal.

Let $\tau_c$ be the chirp length in taps (i.e., the time duration of the chirp is $\tau_c/B$ seconds) and $M$ denote the order of the PPM modulation, or equivalently, the number of bits per chirp. The realized wireless communication with TRDMA for transmitting individual data at two closely-spaced positions $z_1$ and $z_2$ was performed as follows:
\begin{itemize}
    \item Each estimated baseband CIR corresponding to each $z_i$ position, with $i=1$ and $2$, was flipped in time and convoluted with a chirp.
    \item Each $n_i$ bits intended for each $i$-th receiving antenna were converted to an integer $m_i$ that represented the corresponding position of the maximum amplitude. Then, the previously precoded signals were circularly rotated by $m_i$ taps, and then, were transmitted after normalization.
    \item At the side of the receivers, each of them estimated the position of the amplitude peak in the received signal in order to decode the information. To perform this estimation, a simple power detector was used. It is noted that this reception mode is very simple, since it does not require phase detection neither synchronization between the transmit and each receive antennas. 
\end{itemize}

We have evaluated the transmission for different data rates that correspond to different modulation orders and chirp lengths. The data rate for the $2$ users was calculated as follows:
\begin{equation}
    \text{Bit Rate}=2\frac{B M}{\tau_c}.
\end{equation}
Table~\ref{bitrate} includes the different configurations of chirp lengths and orders of modulation that have been used in the experiment, as well as their corresponding bit rates.
\begin{figure}
\centering
\includegraphics[width=1 \linewidth]{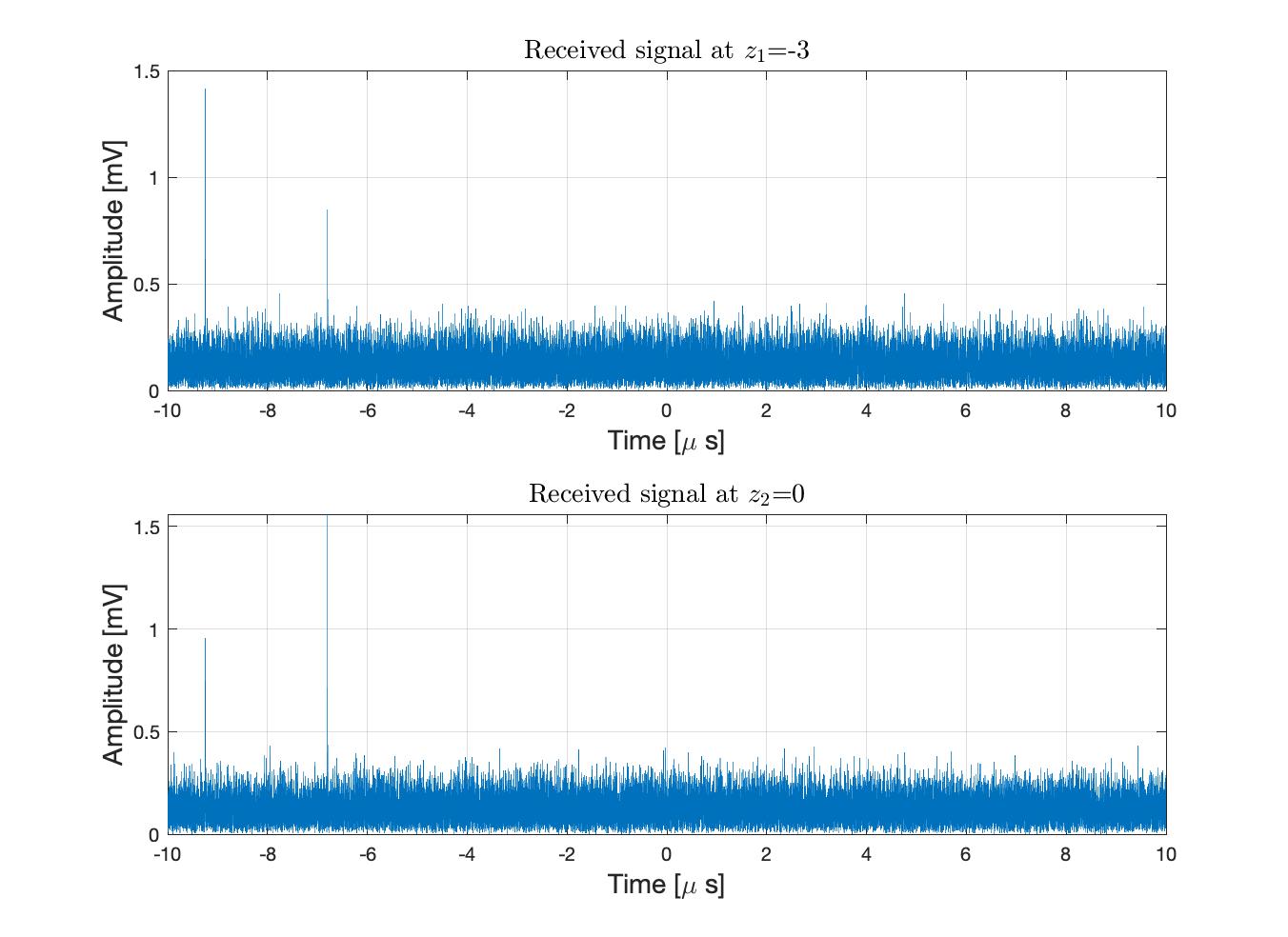}
\caption{Received signals at the positions $z_1=-3$ (up) and $z_2=0$ (bottom) when applying TRDMA with PPM.}
\label{RxTRDMA}
\end{figure}
\begin {table}[t]
\caption{Data rate for different values of $\tau_c$ and $M$.}
\label{bitrate}
\begin{center}
\begin{tabular}{|c|c|c|}
\hline
Chirp length $\tau_c$ [taps]&$M$ per chirp [bits]& Bit Rate [Mbps]\\
\hline
 70&6 &343 \\
\hline
300&8&107 \\
\hline 
1100&10&36.4 \\
\hline 
2200&10&18 \\
\hline
4000&10&10 \\
\hline
40000 &10& 1 \\
\hline

\end{tabular}

\end{center}
\end {table}
Figure~\ref{RxTRDMA} shows the received signals at the two receive positions $z_1=-3$ and $z_2=0$ in mm, when using a chirp of $\tau_c= 40000$ at $2$ GHz bandwidth (i.e., the duration of the chirp was 20 $\mu s$). The order of modulation was $M=10$ and we have set $D=39$ for each receiver. It can be observed that, for each spatial position, the received signal has a maximum value at the temporal position that corresponds to the transmitted PPM symbol, and the second maximum peak represents the IUI from the signal intended for the other position.

\subsection{Signal-to-Interference-plus-Noise Ratio (SINR) Results}
\begin{figure}
\centering
\includegraphics[width=1 \linewidth]{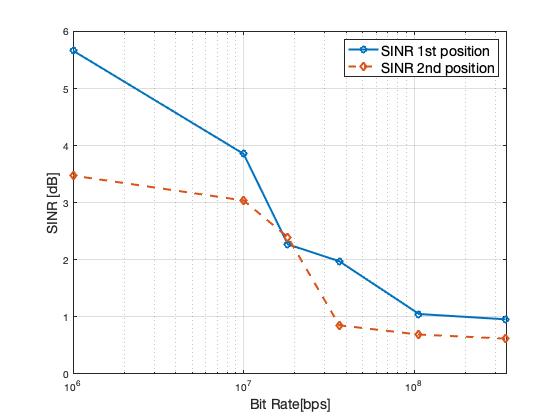}
\caption{SINR performance in dB at the positions $z_1=-2.7$ (1st) and $z_2=-1.8$ (2nd) versus the bit rate in bps.}
\label{SINR}
\end{figure}

We have considered the two receiver positions $z_1=-2.7$ mm and $z_2=-1.8$ mm, and evaluated the SINR of the received signal at each of them, which are depicted in Fig$.$~\ref{SINR} functions of the bit rate, as calculated in Table~\ref{bitrate}. The experiments were done offline (i.e., non-real time measurements), and the CIR estimations were collected at the desktop computer that was connected with both the transmitter and the receiver positions. As shown in the figure, the SINR degrades as the bit rate increases, and this happens as a consequence of the following two factors: \textit{i}) The length of the chirp decreases with a higher bit rate, and in parallel, the energy transmitted per symbol is reduced. \textit{ii}) The rate backoff is reduced in cases of saturation of the order of modulation for a certain chirp length (e.g., when the chirp length is $\tau_c=1100$ taps and the order of modulation is $M=10$ bits per symbol, we have one position per tap). In addition, it is observed that, even for the highest considered data rate, the SINR remains greater than $0$ dB. This verifies the TR capability for multi-access transmissions in the subTHz frequency band. In future work, an experiment with real CIR estimation as well as disconnected transmitter and receivers needs be performed. To keep the receivers simple, the CIR estimation protocol should take into account that the reception is incoherent, hence, no complex data signals should be generated. 

\section{Conclusion}
In this paper, we investigated the capability of TR-based precoding for realizing multi-access transmission in the subTHz frequency band. A novel experimental setup, including a THz link inside a waveguide and operating at the carrier frequency $273.6$ GHz with $2$ GHz transmission bandwidth, was designed and used to first estimate the wireless channel and then perform TRDMA. In particular, estimation of the channel impulse responses and TRDMA at two different receive positions separated by $3$ mm was performed. The presented results validated the spatiotemporal capability of TR for multi-user communications in subTHz frequencies, and showcased data communication with simple PPM detection exceeding rates of $340$ Mbps. In the future, we intend to consider reconfigurable intelligent surfaces \cite{alexandg_2021} for reprogrammable rich scattering conditions, enabling TR-based multi-user wireless communications at the THz frequency band.